\documentclass[]{spie}  
\usepackage[]{amssymb,graphicx}

\def\arcdeg{\hbox{$^\circ$}}
\def\arcsec{\hbox{$^{\prime\prime}$}}
\def\citep{\cite}
\def\citet{\cite}
\newcommand{\bq}{\begin{equation}}
\newcommand{\eq}{\end{equation}}

\def\apj{ApJ}			
\def\apjl{ApJ}		
\def\ao{Appl.~Opt.}		
\def\aaps{A\&AS}		
\def\mnras{MNRAS}		

\title{The Submillimeter Array Polarimeter}

\author{Daniel P. Marrone\supit{a,b}, Ramprasad Rao\supit{c}
\skiplinehalf
\supit{a}Jansky Fellow, National Radio Astronomy Observatory, USA \\
\supit{b}University of Chicago, KICP, Chicago, IL, USA \\
\supit{c}SMA Project, Academia Sinica Institute of Astronomy \&
Astrophysics, Taipei, Taiwan
}

\authorinfo{Further author information: (Send correspondence to
D.P.M.)\\D.P.M.: E-mail: dmarrone@oddjob.uchicago.edu, Telephone: 1
773 834 9785\\  R.R.: E-mail: rrao@sma.hawaii.edu, Telephone: 1 808
961 2935}

\begin{document} 
  \maketitle 

\begin{abstract}
We describe the Submillimeter Array (SMA) Polarimeter, a polarization
converter and feed multiplexer installed on the SMA. The polarimeter
uses narrow-band quarter-wave plates to generate circular polarization
sensitivity from the linearly-polarized SMA feeds. The wave plates are
mounted in rotation stages under computer control so that the
polarization handedness of each antenna is rapidly
selectable. Positioning of the wave plates is found to be highly
repeatable, better than 0.2 degrees. Although only a single
polarization is detected at any time, all four cross correlations of
left- and right-circular polarization are efficiently sampled on each
baseline through coordinated switching of the antenna polarizations in
Walsh function patterns. The initial set of anti-reflection coated
quartz and sapphire wave plates allows polarimetry near 345 GHz; these
plates have been have been used in observations between 325 and 350
GHz. The frequency-dependent cross-polarization of each antenna,
largely due to the variation with frequency of the retardation phase
of the single-element wave plates, can be measured precisely through
observations of bright point sources. Such measurements indicate that
the cross-polarization of each antenna is a few percent or smaller and
stable, consistent with the expected frequency dependence and very
small alignment errors. The polarimeter is now available for general
use as a facility instrument of the SMA.
\end{abstract}


\section{INTRODUCTION}
\label{sec:intro}
As the first dedicated submillimeter interferometer, the Submillimeter
Array (SMA) provides unique capabilities for detailed study of the
dusty sites of star formation. In particular, measurements of
polarized dust emission, which traces the structure of the magnetic
field in the plane of the sky ($B_\perp$), benefit greatly from higher
frequencies and finer angular resolution. The flux density from
optically thin warm dust emission rises like $\nu^{3-4}$, multiplying
the emission by factors of several between the 230~GHz band of
millimeter interferometers such as PdBI and CARMA and the 345~GHz band
of the SMA. Furthermore, structure is expected in the magnetic field
on scales as small as $\sim$100~AU in star forming regions,
corresponding to $\sim$1\arcsec\ in even the nearest objects. Previous
observations of polarized dust emission have generally been made at
lower frequency and few arcsecond resolution or higher frequencies
with no better than 10\arcsec\ resolution. The SMA provides the first
opportunity to examine polarization at both high frequencies and high
angular resolution. Illuminating the important and poorly understood
role of the magnetic field in the star formation process provides a
strong motivation for adding a polarimetric capability to the SMA. In
addition, there are other targets, such as the polarization signatures
of low-luminosity accretion\cite{MarroneE06}, that are accessible only
near submillimeter frequencies.

The SMA was designed to allow polarimetry, but requires additional
hardware to enable these observations. Most importantly, the SMA can
measure only a single linear polarization at a given frequency,
although this will change for some receivers in future
upgrades. Therefore, modulation of the polarization sensitivity of
each antenna is required in order to recover complete polarization
information. Moreover, the interferometric measurement of linear
polarization is best made with circularly-polarized receivers (see,
e.g., ref.~\citenum{URP2}), so a linear-to-circular polarization
conversion is desirable. The instrument described here provides the
polarization conversion and rapid modulation that is required for
these observations. We outline the instrument design, control system,
calibration procedures and results, and a provide examples of the
science enabled by the polarimeter. 

\section{POLARIMETER INSTRUMENT}
\label{sec:inst}

\subsection{Telescope Optics}
The Submillimeter Array is composed of eight 6-meter diameter antennas
on alt-az mounts. In these Cassegrain telescopes the receivers are
mounted in a fixed cryostat at a Nasmyth focus. The relay optics
between the secondary mirror and the cryostat consist of four
mirrors\footnote{The optical system is described in the SMA Project
Book, http://sma-www.cfa.harvard.edu/private/eng\_pool/table.html};
two of these are curved and serve, with the lenses that are specific
to each receiver band, to image the receiver feedhorn apertures onto
the secondary mirror. The optics also generate an intermediate image
of the feeds at an accessible location between the curved mirrors that
is intended to accommodate calibration loads and polarizing
elements. The polarimeter hardware is designed for installation at
this location for polarimetric observations and easy removal to return
to normal SMA operation.

\subsection{Quarter-wave Plates}
The linear-to-circular polarization conversion that is central to
polarimetry with the SMA is achieved with a quarter-wave plate
(QWP). A QWP is an anisotropic optical element that introduces a delay
of one quarter wavelength between orthogonal linear
polarizations. When the QWP axes are oriented at 45\arcdeg\ to
incident linear polarization, the polarization component along the
``slow'' axis of the wave plate is delayed and pure circular
polarization is produced. By rotating the QWP by 90\arcdeg\ the
handedness of the circular polarization can be reversed.

Wave plates are ubiquitous in polarimetric experiments. Half-wave
plates rotate linear polarization and are often used to modulate the
polarization sensitivity of a detector of fixed polarization. Examples
include the polarimeters for the submillimeter cameras on the
JCMT\cite{SCUBApol} and CSO\cite{LiE08}, and various CMB polarization
experiments\cite{MAXIPOL,EBEX}. Quarter-wave plates have
also been employed to generate circular polarization for interferometers
with linear feeds, including the Berkeley-Illinois-Maryland
Array\cite{Rao99} and the Owens Valley Radio Observatory millimeter
array\cite{AkesonE96}. 

In the simplest design, half- or quarter-wave plates are single
elements of birefringent material tuned in thickness to provide the
appropriate path delay. However, this difference in propagation
distance provides the desired phase delay at just one frequency, with
a phase error that increases linearly away from this frequency. More
complicated designs incorporating multiple elements can provide a
nearly fixed phase delay over a large
bandwidth\cite{Pancharatnam55,Title75}. These designs are less
frequently used but may become common as a result of the bandwidth
requirements of bolometer-based CMB polarization
experiments\cite{HananyE05}. In the case of the SMA, such a broadband
design could allow polarimetry across the full RF bandwidth of a given
receiver band, or possibly all bands at once for very complicated
designs\cite{MassonGallot06}. Nevertheless, SMA science goals do not
justify the added complication and loss of these designs, as there are
few spectral lines that are interesting for polarimetry at SMA
sensitivity and little to be gained by continuum polarimetry at two
frequencies within a single receiver band. Instead, the 345~GHz QWPs
of the SMA polarimeter use the single-element design, designed for an
optimal frequency near the $J=3\rightarrow2$ line of CO
(345.8~GHz). This nearly bisects the most interesting frequency range
in this band, between the atmospheric cutoff near 360~GHz and the
masing\cite{MentenE90} water line at 325~GHz, minimizing the
retardation error at both ends of the range.

Wave plates can be made from materials with natural or manufactured
birefringence. The SMA wave plates are made from low-loss naturally
anisotropic crystals, either quartz or sapphire
(Table~\ref{tab:plates}). To minimize reflections from the QWPs, which
degrade the receiver sensitivity and can generate standing waves in
the optical path, anti-reflection (AR) coatings have been
applied. The ideal (monochromatic) AR coating is a quarter-wave thick
with an index of refraction that is the geometric mean of the indices
of the materials on either side of it. For quartz in air this is
roughly 1.46, which is fairly close to the index of low-density
polyethylene (LDPE), $n_{LDPE} = 1.514$\cite{Lamb96}. The larger
index of sapphire is not particularly well matched by common polymer
materials and for simplicity was also coated with LDPE. The expected
reflectivity of the sapphire plates over the range of operating
frequencies is 0$-$7\%, much larger than the expected $<1$\%
absorption in the sapphire. Measurements of the quartz wave plates near
the optimum frequency indicate that they are 98\% transmissive,
resulting in a 9\% increase in the system temperature under typical
operating conditions.

\begin{table}[h]
\caption{Properties of the 345~GHz quarter-wave plates of the SMA
polarimeter.}
\label{tab:plates}
\begin{center}       
\begin{tabular}{lcc} 
\hline
 & Quartz Plates & Sapphire Plates\\
\hline
Quantity & 7 & 1 \\
Material & X-cut quartz & A-cut Hemlux sapphire \\
Indices of Refraction & & \\
\hspace{0.5cm}Ordinary & 2.106\cite{BirchE94} & 3.064\cite{Afsar87} \\
\hspace{0.5cm}Extraordinary & 2.154\cite{BirchE94} & 3.404\cite{Afsar87} \\
Thickness (mm) & 4.566$\pm$0.003 & 0.640 \\
Central Frequency (GHz) & & \\
\hspace{0.5cm}Nominal & 342.0 & 345.5 \\
\hspace{0.5cm}Measured & 347.7$\pm$1.5 & 339.5 \\
Anti-Reflection Coating & & \\
\hspace{0.5cm}Material & LDPE & LDPE \\
\hspace{0.5cm}Mean Thickness ($\mu$m) & 140$-$160 & 133 \\
\hspace{0.5cm}Peak-to-Peak Variation ($\mu$m) & 3$-$14 & 10 \\
\hline 
\end{tabular}
\end{center}
\end{table} 

\subsection{Positioning Hardware}
\begin{figure}[t]
   \begin{center}
   \begin{tabular}{cc}
   \includegraphics[height=7.5cm]{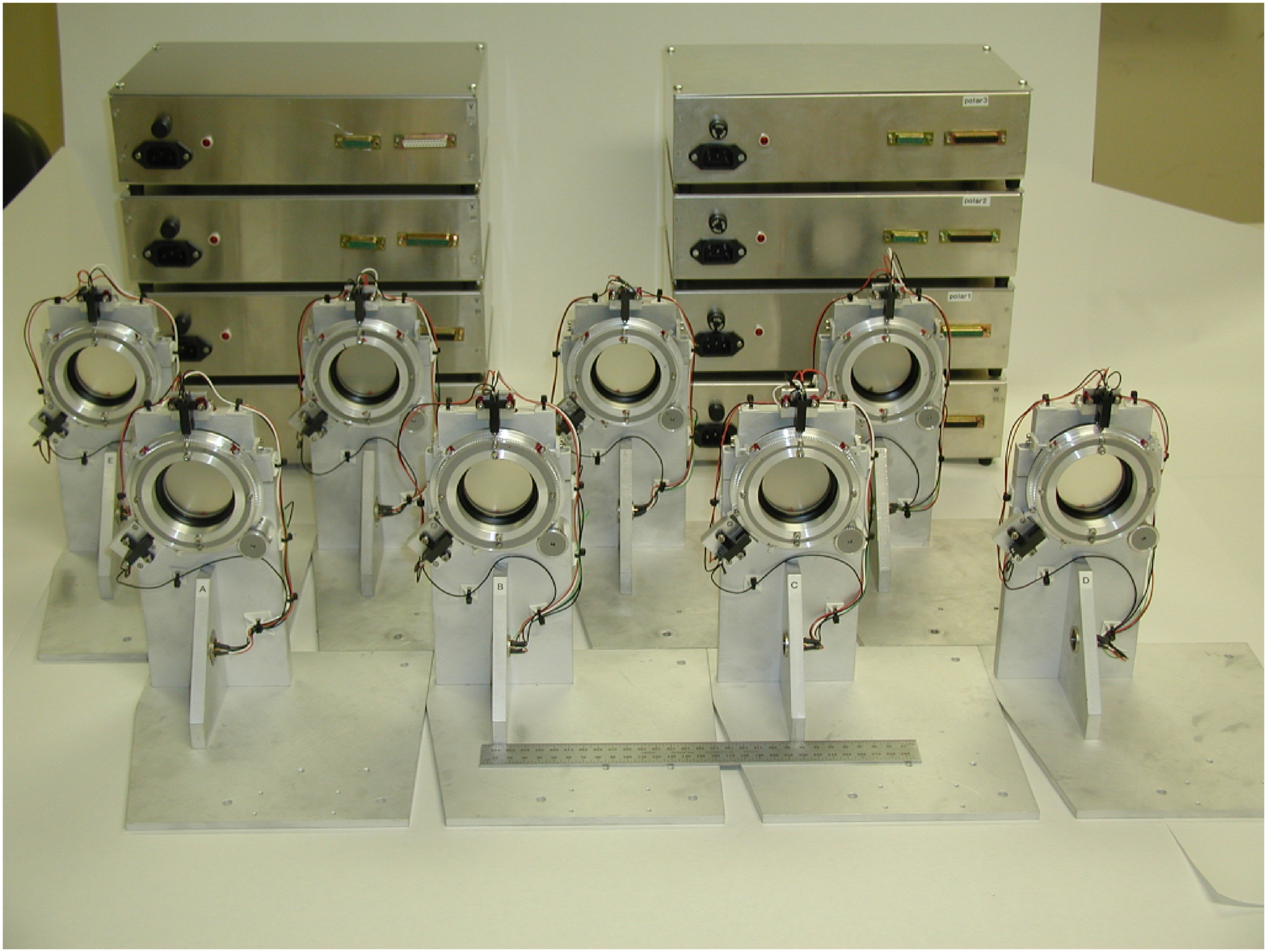} &
   \includegraphics[height=7.5cm]{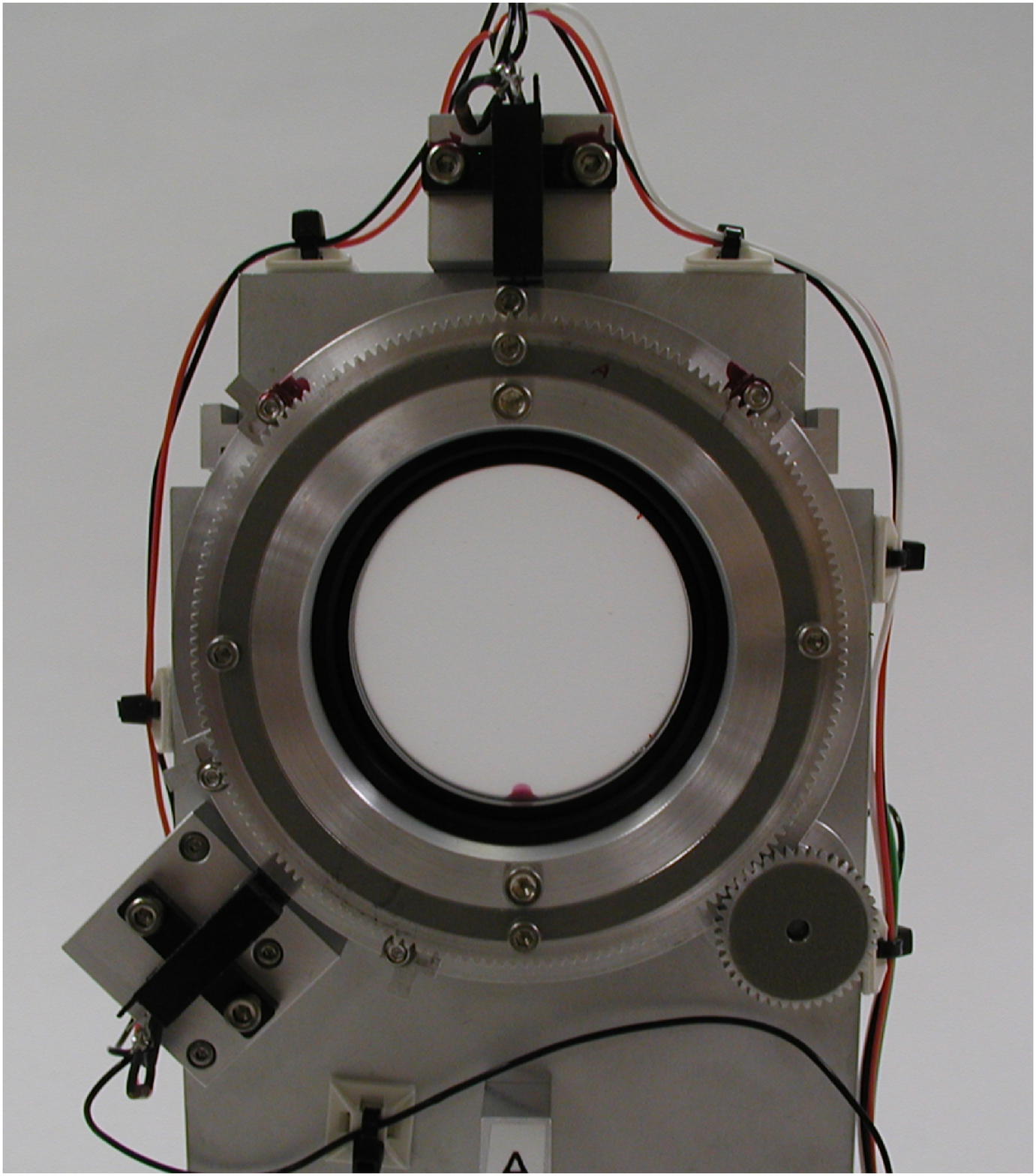} 
   \end{tabular}
   \end{center}
   \caption[] 
   { \label{fig:polarimeter} {\it Left}: The wave plate positioners
   and control computers of the SMA polarimeter. {\it Right}: Closeup
   of the rotation stage and QWP. Optical brakes are visible at the
   top and the bottom left. The adjustable flags that determine the
   wave plate positions are visible on the circumference of the
   rotation stage at $\pm45^\circ$ from the optical brake at the top
   and at $\pm22.5^\circ$ from the brake at the bottom left.}
\end{figure}

In order to convert the SMA to circular polarization sensitivity the
QWPs must be repeatably aligned to the feed polarization and be
rotatable to select LCP or RCP. This is achieved with a
computer-controlled mount and rotation system, the wave plate
positioning hardware (Fig.~\ref{fig:polarimeter}). The QWPs are held
in a rotation stage driven by a gear and DC motor. There is no encoder
to monitor the position of the motor or rotation stage, rather, the
wave plate positions are determined by two optical brakes
(opto-interrupters) and four adjustable positioning flags on the
circumference of the rotation stage (Fig.~\ref{fig:polarimeter},
right). These flags determine four angular positions for the wave
plate, two of which set the $\pm45^\circ$ position angles that convert
the feed polarization to left- and right-circular polarization. The
other two flags are monitored by a second optical brake and could be
used to set positions $\pm22.5^\circ$, allowing half-wave plates to
switch a feed between two linear polarizations 90\arcdeg\ apart.

Stability of the polarization properties of the SMA requires precisely
repeatable positioning of the QWPs. Any unnoticed error in
positioning, such as that due to backlash or lost encoder counts,
would introduce a change in the instrumental polarization and corrupt
the calibration. We have used optical brakes because they are immune
to backlash in the motor and do not rely on repeated contact, as a
hard stop would. To accurately recover the intended stopping point for
each orientation the edge of the positioning flag is found via a
two-step process. The coarse alignment is derived from a slew at the
maximum motor speed that ends when the flag interrupts the beam. Due
to differences in the friction between wave plate mounts, the position
determined in this way can be inaccurate by a degree or more and might
be expected to change with temperature as the pressure on the bearing
changes. Much greater accuracy is obtained in the second step, where
the motor makes rapid pulsed movements until the correct edge of the
flag (depending on the initial and final positions) is driven out of
the opto-interrupter beam. Measurements of the positional
repeatability achieved through this procedure found it to be better
than the measurement precision, around 0.2\arcdeg, which is more than
adequate given the astronomical calibration limitations.

The wave plate positioning hardware operates and interfaces with the
rest of the array through dedicated control computers. Each wave plate
mount is controlled by a compact (PC-104 form-factor) disk-less
computer that is accessed via a remote procedure call (RPC)
application. The RPC server controls the plate movements, reports
failures in acquiring positions, and can re-initialize the rotation
stage to re-establish the location should it be lost.

The SMA QWPs must switch rapidly between the left- and right-circular
polarization states to limit the observing time lost to switching. The
motor and gear ratio have been chosen to ensure that the switch
between the widely separated LCP and RCP positions can be achieved in
less than 2~seconds in either direction. Upon initial deployment the
positioners were found to switch in 1.5$-$1.9 seconds. Four years
later the switching time has degraded very little, to 1.6$-$2.0
seconds in a recent measurement.

\subsection{Polarimetric Observations}
At present, most SMA antennas have only a single 345~GHz feed and
therefore cannot sample both polarizations at the same frequency
simultaneously. To recover the full polarization information we use
the wave plates to switch the antennas between LCP and RCP in a
coordinated temporal sequence. The switching patterns of the antennas
are described by orthogonal two-state sequences known as Walsh
functions. Well-chosen patterns will sample all four polarization
combinations four times each on every baseline in a 16-step
cycle. Averaging the visibilities over the cycle time results in
quasi-simultaneous measurements of all polarizations. Under typical
observing conditions this cycle is completed in 4$-$6~minutes.

The polarization switching results in significantly poorer sensitivity
than that achieved in normal SMA observations. A significant portion
of the degradation is caused by the time lost to switching
polarization states and the process of stopping and restarting the
correlator. For the 15-second integrations normally used, an
additional 5 seconds per integration are lost to this overhead. Longer
integrations would reduce the fractional impact of the overhead but
risk decoherence when averaging over the polarization cycle. The most
important cause of sensitivity loss is the time-multiplexing of the
polarization sensitivity. On any baseline only half of the
integrations in the polarization switching cycle are sensitive to
Stokes I (LL and RR correlations) and half sensitive to Stokes Q and U
(the linear polarization, LR and RL correlations). Combining these two
effects, the integration time for the total intensity or linear
polarization is approximately 3/8 of that for normal SMA
observations. Furthermore, the QWPs raise system temperature by
$\sim$10\%, as discussed above, for a total increase in the rms noise
of approximately 1.8$\times$.

\section{CALIBRATION}
\label{sec:calib}
\subsection{Instrumental Polarization}

Proper determination of the polarization of an astronomical source
requires calibration of the instrumental polarization. The response of
the SMA antennas, including the feeds and wave plates, can be modeled
by a term representing the sensitivity to the desired polarization
plus a fractional sensitivity to the orthogonal polarization, a
complex ``leakage'' term\cite{URP2}. To lowest order, the leakage
mixes the total intensity (Stokes I) with the linear polarization (Q
and U). For typical submillimeter sources the astronomical linear
polarization is no more than ten percent, while the leakage terms
are often of similar magnitude, so these contributions must be removed
precisely in order to determine the polarization of the target source.
Fortunately, the instrumental and astronomical polarizations are
separable for alt-azimuth mounted telescopes and sources of known polarization
structure. Telescopes with non-equatorial mounts, such as the SMA
antennas, see an apparent rotation of the target source as it moves
across the sky, which distinguishes source and antenna
polarization. Therefore, observations of a bright point source over a
large range of parallactic angles allow determination of the leakages
and are included in most SMA polarimetry tracks. The frequency
coverage of the leakage measurements has been determined by the
science goals of the polarimetric projects, resulting in very
redundant data at few frequencies. This type of sampling provides very
good measurements of stability and reproducibility in the leakage
determinations, as described below. The spectral properties of the
leakages are more poorly determined, but can also be examined from
these data (Sec.~\ref{ssec:lfreq}).

\subsection{Stability and Measurement Precision}
\label{ssec:stab}
\begin{figure}
   \begin{center}
   \begin{tabular}{c}
   \includegraphics[width=5in]{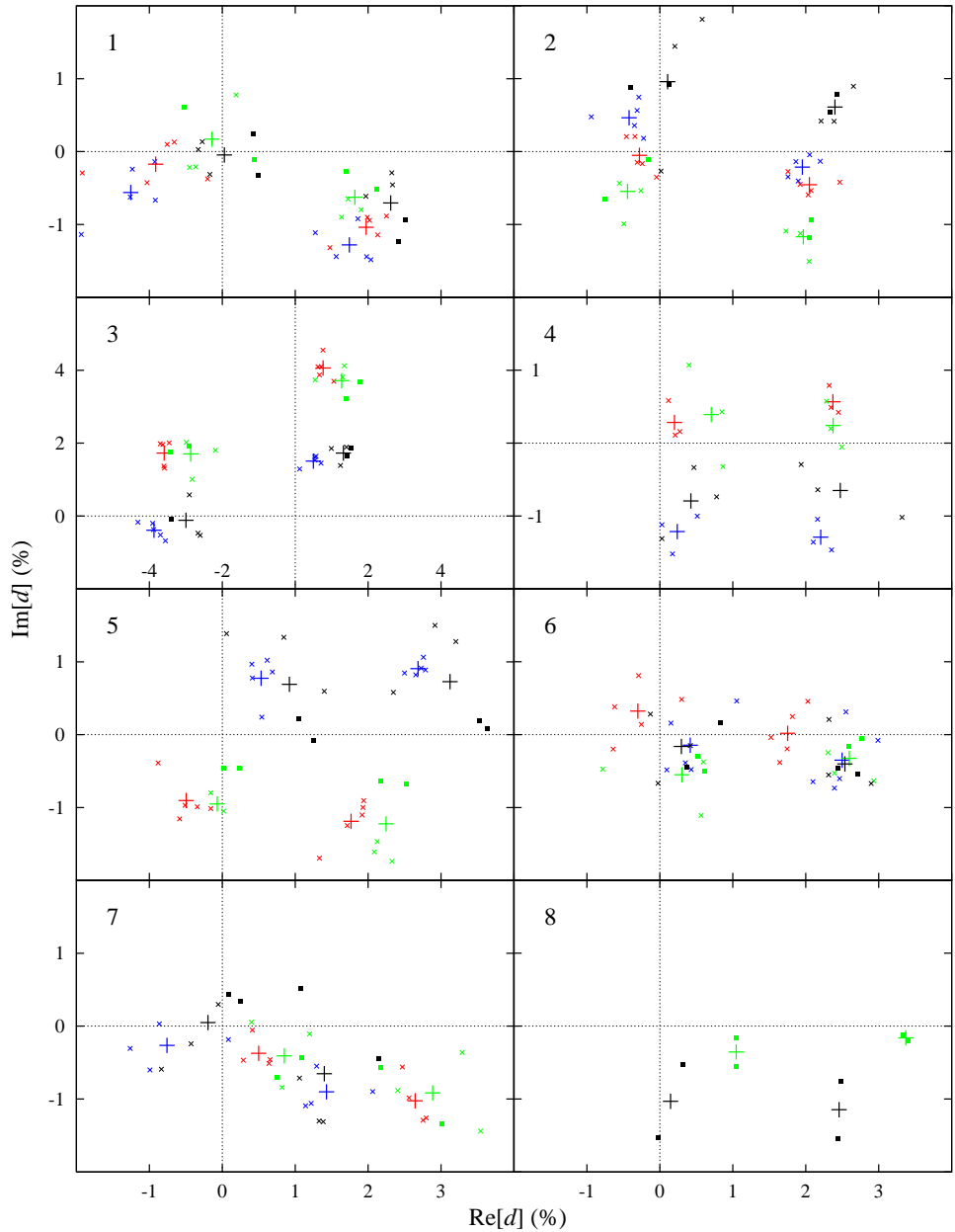} 
   \end{tabular}
   \end{center}
   \caption[] 
   { \label{fig:leaks} Leakages measured in 2005 and 2006. Each plot
   represents a single antenna, identified at the upper left. Leakages
   are plotted in the complex plane, with with equal linear scales in
   the real and imaginary axes. The red and blue points represent
   $d_R$ and $d_L$, respectively, at 338.0~GHz (points at right in
   each plot) and 348.0~GHz (points at left). Green and black points
   represent $d_R$ and $d_L$, respectively, at 336.5~GHz (right) and
   346.5~GHz (left). The mean value of $d_R$ ($d_L$) for each
   frequency is represented by a large red/green (blue/black)
   $+$. Individual measurements from 2005 ($\times$) and 2006
   ($\scriptstyle{\blacksquare}$) are marked with smaller symbols. The
   subplot for antenna 3, which hosted the sapphire plate, has a
   linear scale twice as large as those for the other antennas, as
   shown by its labels. }
\end{figure}

Fig.~\ref{fig:leaks} shows the real and imaginary components of the
leakages measured for each antenna at four frequencies that have been
repeated several times (two array tunings with two sidebands 10~GHz
apart for each tuning). The leakages were measured over a period of a
few months in 2005 and 2006, allowing investigation of the
repeatability of the measurement and the stability of the
system. Measurements at the same frequency and for the same
polarization state (leakage of RCP into LCP, $d_R$, or LCP into RCP,
$d_L$) form tight groups, even between years. The leakages appear to
be as stable on long (one year) timescales as they are across periods
of weeks. The spread in values in the Re[$d$] direction is due to the
linear variation of the QWP retardation with frequency, this is
discussed further below.

Taking all of the measurements together, the average scatter in the
leakage measurements is 0.4\%. For most of these measurements, made at
a time when the calibration source (3C454.3) was undergoing a large
flare and was exceptionally bright, the measurement uncertainty was
0.1$-$0.15\%, smaller than the scatter. If the scatter represents a
random variation in the leakages that is uncorrelated between
antennas, the expected false polarization signal introduced into an
snapshot polarization observation would be 0.15\%. In a long synthesis
this is further reduced by the parallactic angle rotation of the
source polarization. In the other extreme, if the leakage changes in
any given observation are fully correlated in a way that directly
converts $I$ to $Q$ or $U$, as much as the full leakage error may be
introduced into a snapshot polarization measurement. Just as in the
case of random errors, parallactic angle rotation will reduce the
average polarization error across the whole track. However, there are
few systematic errors that should operate to introduce correlated
leakage errors and these data do not show such correlation, so this
worst case is unlikely to be realized.

\subsection{Frequency Dependence}
\label{ssec:lfreq}
To first order the leakage ($d$) depends on the frequency offset
($\delta=\left[\nu-\nu_0\right]/\nu_0$), orientation error ($\theta$),
and difference in the field reflection coefficient along the two QWP
axes ($\epsilon$) as\cite{Marrone06}
\bq
d \simeq -\frac{\pi\delta\left(\nu\right)}{4}+
\left(-\theta+\frac{\epsilon\left(\nu\right)}{2}\right)i .
\eq
Fitting the variation of the real component of the leakage with
frequency, visible in Fig.~\ref{fig:leaks} provides a determination of
the optimum frequency of the plates. For the seven quartz plates this
is 347.7~GHz with a scatter of 1.5~GHz, while the sapphire plate is
tuned to 339.5~GHz. Differences between the nominal and observed
central frequencies are largely due to the uncertainty in the
literature values for the birefringence ($n_e-n_o$) of the QWP
materials (10\% for quartz). The slope of the variation with frequency
should also be inversely proportional to the optimum frequency; the
value of $\nu_0$ derived in this way is poorly constrained but agrees
with the intercept results. The imaginary components provide a measure
of the alignment error in the absence of the $\epsilon$ term, but
unfortunately the scatter in coating thickness ensures that there is
non-negligible reflectivity in some of the plates. From other
techniques, the alignment errors are found to be 0.3\arcdeg\ or less.

\section{SAMPLE SCIENTIFIC RESULTS}
\label{sec:science}
\begin{figure}
   \begin{center}
   \begin{tabular}{c}
   \includegraphics[height=10cm,angle=270]{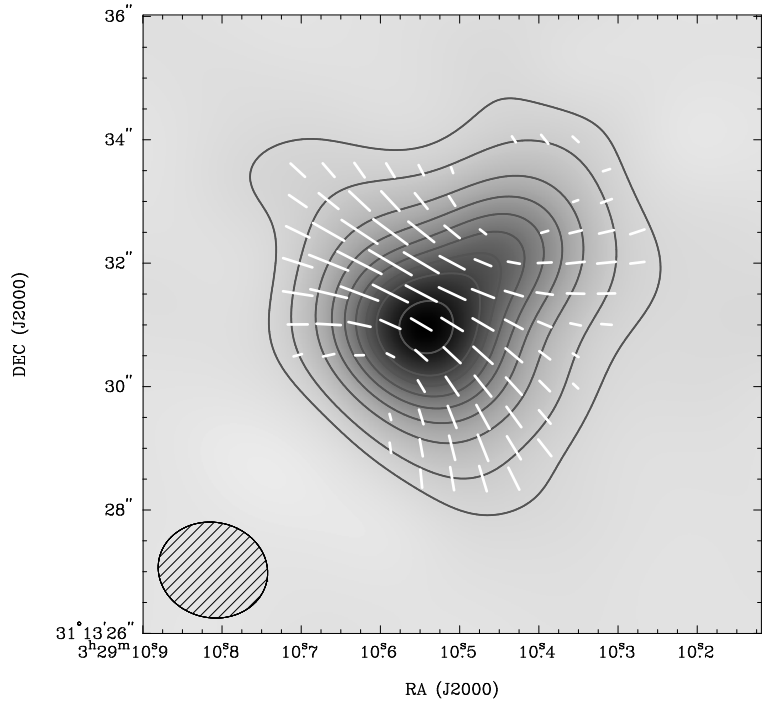}
   \end{tabular}
   \end{center}
   \caption[] 
   { \label{fig:ngc1333} Polarization in the protostar NGC 1333 IRAS
   4A, as measured by the SMA polarimeter. Continuum emission is shown
   in grayscale and contours. The magnetic field direction (orthogonal
   to the polarization direction) is shown with vectors, their length
   is proportional to the polarized flux. This observation provided
   the first clear evidence of an hourglass-shaped magnetic field on
   the scales expected for isolated star formation\cite{GirartE06}.}
\end{figure}

The SMA polarimeter was installed on the array in April, 2004 and
became available for general use as a facility instrument in November,
2005. The first scientific result from the instrument\cite{MarroneE06}
was the measurement of the 345~GHz linear polarization of the
supermassive black hole at the center of our Galaxy,
Sagittarius~A*. Thanks to the angular resolution and sensitivity of
the SMA, the emission from Sgr~A* was cleanly separated from the
surrounding dust emission and intraday variations in the polarization
were detected for the first time. Many subsequent observations have
also investigated the magnetic structure in star-forming
regions. Fig.~\ref{fig:ngc1333} is an example\cite{GirartE06} of this
type of observation in the low-mass protostellar system NGC~1333
IRAS4A. The structure in the field on arcsecond scales matches very
well with theoretical predictions of the warping of an initially
uniform magnetic field due to the collapse of the magnetized
cloud. However, previous observations had lacked the sensitivity and
angular resolution to confidently detect this ``hourglass'' shape. 

\section{CONCLUSIONS AND FUTURE IMPROVEMENTS}
\label{sec:concl}
We have presented the 345~GHz polarimeter instrument for the
Submillimeter Array. This instrument now provides a fully-supported
polarization observing mode for the SMA, used in roughly 10\% of
observations with the array. Several upgrades to the polarimeter are
planned or have been completed since its deployment. As part of a new
calibration system, the installation and removal of the wave plates
was automated in 2007. Additional wave plates for 230/690~GHz and
240/400~GHz have been added; commissioning of these plates is
ongoing. Over the coming year, the SMA will gain a new 340-420~GHz
receiver that will provide dual-polarization observations near
345~GHz. This obviates the need for rapid switching of the antenna
polarization state, although some polarization modulation is still
desirable to prevent errors in the orientation of the derived
polarization due to inter-receiver phase drift. The decrease in the
switching overhead and the instantaneous measurement of all four
polarized cross correlations will improve the sensitivity by a factor
of more than $\sqrt{5}$ for this band, greatly increasing the number
of sources accessible to the array.

\acknowledgments
We are grateful to the staff of the SMA and the SAO submillimeter
receiver lab for their assistance during the design and testing of the
polarimeter. We thank Ken Young for his tireless work adapting the
correlator software to accommodate this new observing mode. We also
thank the director of the SMA, Ray Blundell, and the past director,
Jim Moran, for their support for this instrument.


\end{document}